# The Hermeneutic Turn of AI: Are Machines Capable of Interpreting?

Rémy Demichelis, PhD,
Visiting Researcher,
University of Turin

**Abstract**

This article aims to demonstrate how the approach to computing is being disrupted by deep learning (artificial neural networks), not only in terms of techniques but also in our interactions with machines. It also addresses the philosophical tradition of hermeneutics (Don Ihde, Wilhelm Dilthey) to highlight a parallel with this movement and to demystify the idea of human-like AI.

## Introduction

The notion of interpretation is increasingly present in the world of artificial intelligence (AI). For humans, it involves interpreting algorithms that are difficult to explain mathematically. For machines, the challenge is to interpret data to draw conclusions. They must now also interpret brief instructions in natural language: this is the operational principle of ChatGPT and other chatbots grounded on generative AI which interacts verbally with unsettling fluidity. We can thus speak of a true interpretive turning point in AI.

The art of interpretation, however, has been known for centuries under the term "hermeneutics". It initially applied to the reading of poets or sacred texts before evolving into a philosophical current to signify that interpretation is at the foundation of understanding, or even that it represents the necessary activity of who we are (Gadamer, 2004-1960; Heidegger, 1962-1927; Nietzsche, 1954-1886/1887, 2003-1887). Our access to the world is indeed always influenced by certain tones that are not neutral but carry a cultural charge. Does the resemblance stop however at the mere use of the term interpretation? In other words, is AI doing hermeneutics? Should we use the art of interpretation to understand machines? Or is it both?

## 1. Direct Dialogue with the Machine in Our Language

The event that completes the interpretive turn of AI is undoubtedly the launch of ChatGPT in November 2022. The essential innovation of large language models, like its own, is that the machine is required to interpret human instructions more than ever before. The user inputs a prompt to request what they want, and then the system provides a response, whether it's text, an image, or spoken output.

We no longer address the machine in computer language, in code, but in natural language or what is called unstructured data. Certainly, the "hallucinations" (errors of the machine in the form of plausible but delusional statements) are countless, and the results can still be improved, but something is happening. Interpretation, an activity we long believed was reserved for humans, is now being undertaken by digital tools.

Computing has already long been an object of interpretation, since science has increasingly relied on digital instruments and imaging techniques (medical, nanometric, spectroscopic, etc.). The American philosopher Don Ihde, who passed away this year, noticed this early on, first in his work "Technology and the Lifeworld" (Ihde, 1990).

Unfortunately, it is only after his death that his relevance seems to leap out at us. "All imagery calls for interpretation", he wrote (Ihde, 2021, Chapter 1). He goes on to explain that imagery is "materially technological in its instrumental embodiment" because it requires the use of a sophisticated tool to produce it, reveal the image, and thus the object being studied.

> [His] idea for a material hermeneutics is closely tied to the 20$^{th}$-21$^{st}$ shift to micro-processing imaging technologies that transform science practice and evidence production [...] These technologies helped enhance the necessity for interpretation or hermeneutics practices in natural science. (Ihde, 2021, Chapter 4)

For Ihde, what characterizes this "necessity for interpretation" is that we are no longer being in a direct relationship with things. We must go through instruments or images in such a way that we construct the object through the medium, like a camera or a scientific measuring instrument. Our understanding of the object is then inseparable from the medium without which we could not know it. The famous photograph of a black hole (Collaboration *et al.*, 2019), which is not exactly a photograph but a construction from data from eight different radio telescopes, provides one of the best illustrations of this.

## 2. The Return of Ambiguity

According to Ihde, the interpretive turn that science has taken tends to bridge the gap between "explanation" and "understanding" (Dilthey, 1989-1883). It is one thing to explain how a castle was built, with what materials or techniques. It is another to understand the reason for its existence, why its builders decided to erect it in a particular place at a certain time. In this latter case (that of understanding), it is necessary to call upon interpretation, in light of historical elements. However, science increasingly veers into *interpretation* in order not to merely *explain* the objects it studies. This marks a rapprochement between the sciences and the humanities (literature, philosophy, history...).

AI further accentuates this rapprochement. Already because the machine is asked to interpret what is given to it, but also because humans must increasingly interpret the results produced by the machine (Lundberg & Lee, 2017; Ribeiro *et al.*, 2016). Ambiguity is taking on a growing role in the world of computing, which, as an heir to mathematics, believed it was preserved from such things. And where there is ambiguity, there is also interpretation. Current AI systems, particularly image analysis or text generation, rely on artificial neural networks. This "deep" learning technique, however, is not easily grasped, even by experts. This is particularly damaging when we realize later that the machine reproduces discriminatory bias (Bernheim & Vincent, 2019; Buolamwini, 2023; Buolamwini & Gebru, 2018; Gebru, 2020; Lowry & Macpherson, 1988; Noble, 2018; O'Neil, 2016).

The AI Act, a regulation on AI recently adopted by the European Union (EU 2024/1689), however, stipulates that so-called "high-risk" systems must undergo thorough analyses (the nature of which remains to be defined). But it is impossible to determine exactly why the software produces a particular result; we can only "interpret" its functioning. While there are today techniques of "explainability" to estimate the weight of each variable, it is indeed the term "interpretability" that should be favored, as they only offer estimates, but no clear and distinct explanation, the kind that mathematics requires to eliminate any ambiguity.

AI even invites us to go beyond quantitative interpretations, since it is important to understand historically how AI models construct their sometimes biased or discriminatory interpretations:

> Even if someone could convince themselves that algorithms sometimes just spit out nonsense, the structure of the nonsense will tend vaguely toward the structure of historical prejudices. (Gebru, 2020)

While interpretability techniques will have their utility, it will also be necessary to analyze the outputs of AI in a more sensitive way, considering that they are also the product of a specific history and society (Kudina, 2023).

## 3. Interpreting to Find Meaning

If AI is indeed capable of interpreting our statements to some extent in order to generate an answer, understanding seems to be nevertheless a phenomenon that goes beyond this. To understand something requires a certain amount of imagination to conceive the object of our knowledge in its multiple and new configurations, to grasp it in a way that is rarely formal but comes through feeling. Some pupils recite their lessons admirably without understanding anything, as they lack the necessary feeling to exclaim: "Got it!" or, as Archimedes is believed to have said, "Eureka!" This feeling is almost impossible to describe, but have you never marveled at suddenly understanding something that had resisted you? If so, you know well what this feeling is, this sensory event of understanding.

And this feeling is fertile, as it can produce interpretation: new connections appear, new configurations, new horizons that feed our imagination. We sometimes say: "that makes sense" and it really does. It makes sense, literally, as I feel an interpretation to be right. This is then an aspect of interpretation that separates our understanding from what machines do, since computer systems are unable to feel. The imagination necessary for the art of interpretation will always be for them an impoverished version of it, an "e-magination" (Romele, 2020).

The interpretation produced by generative AI thus differs from ours in that it is incapable of understanding anything. Nevertheless, it represents a decisive aspect of the interpretive turn that unfolds in various ways in the world of sciences.

The machine interprets our requests in natural language, and we interpret its results or functioning. AI brings hermeneutics back into fashion to the point that we should speak not just of artificial intelligence, but of artificial interpretation.

*This text was originally published in "The Conversation", June 3, 2024 (Demichelis, 2024), under the Creative Commons CC BY-ND 4.0 License. Read the original article.*